\documentclass[aps,prapplied,bibnotes,footinbib,twocolumn]{revtex4-2}

\usepackage{graphicx}
\usepackage{amsmath}
\usepackage{multirow}
\usepackage{braket}
\usepackage{hyperref}
\usepackage{xcolor}

\begin{document}

\title{Bidirectional deep learning of polarization transfer in liquid crystals\\ with application to quantum state preparation}

\author{Dominik Va{\v{s}}inka}
\email{vasinkadominik@seznam.cz}
\affiliation{Department of Optics, Palack\'{y} University, 17. listopadu 12, 77146 Olomouc, Czechia}

\author{Martin Bielak}
\affiliation{Department of Optics, Palack\'{y} University, 17. listopadu 12, 77146 Olomouc, Czechia}

\author{Michal Neset}
\affiliation{Department of Optics, Palack\'{y} University, 17. listopadu 12, 77146 Olomouc, Czechia}

\author{Miroslav Je{\v{z}}ek}
\email{jezek@optics.upol.cz}
\affiliation{Department of Optics, Palack\'{y} University, 17. listopadu 12, 77146 Olomouc, Czechia}

\begin{abstract}
Accurate control of light polarization represents a core building block in polarization metrology, imaging, and optical and quantum communications. Voltage-controlled liquid crystals offer an efficient way of polarization transformation. However, common twisted nematic liquid crystals are notorious for lacking an accurate theoretical model linking control voltages and output polarization. An inverse model, which would predict control voltages required to prepare a target polarization, is even more challenging. Here we report both the direct and inverse models based on deep neural networks, radial basis functions, and linear interpolation. We present an inverse--direct compound model solving the problem of control voltages ambiguity.
We demonstrate one order of magnitude improvement in accuracy using deep learning compared to the radial basis function method and two orders of magnitude improvement compared to the linear interpolation. Errors of the deep neural network model also decrease faster than the other methods with an increasing number of training data. The best direct and inverse models reach the average infidelities of $4 \times 10^{-4}$ and $2 \times 10^{-4}$, respectively, which is the accuracy level not reported yet.
Furthermore, we demonstrate local and remote preparation of an arbitrary single-photon polarization state using the deep learning models. The results will impact the application of twisted-nematic liquid crystals, increasing their control accuracy across the board. The presented bidirectional learning can be used for optimal classical control of complex photonic devices and quantum circuits beyond interpolation.
\end{abstract}

\pacs{}

\maketitle

\section{Introduction}

A vast number of applications require accurate control of the polarization state of light, such as in display technology, ellipsometry, polarization microscopy, and optical communications. The polarization can be manipulated by mechanically adjusting birefringent elements or using electro-optic modulators. The former approach can reach high accuracy but is slow, producing vibrations, and prone to malfunction. The latter one is ultra-fast and vibration-free, however, its accuracy is limited. Free-space electro-optic modulators are bulky and require high voltage drivers, which tend to fluctuate and decrease the overall accuracy even further. Low-voltage integrated modulators show inherent polarization instabilities due to fiber coupling.

Liquid crystals represent a middle ground between the stability of birefringent elements and the response speed of electro-optic modulators. Voltage-controlled nematic liquid crystals are particularly convenient and widely available, as they are commonly used in the display industry.
We distinguish two main types of nematic liquid crystals based on the alignment of the crystals in a device, namely parallel and twisted configurations. The former acts as a polarization retarder and is typically custom-made for specialized applications; the latter is used in displays.
Parallel nematic liquid crystals were utilized as polarization retarders for polarization modulation~\cite{Zhuang1999}, polarization state preparation and tomography~\cite{Adamson2007,Adamson2010}, remote state preparation and imaging~\cite{Peters2005,Defienne2021}, entangled-photons generation~\cite{Lohrmann2019, Villar2020}, and implementation of quantum channels~\cite{Fisher2012,Orieux2013} and quantum communication protocols~\cite{Naik2000,Barreiro2008}.
Recently, fully reconfigurable topological photonic devices were proposed employing nematic liquid crystals \cite{Abbaszadeh2021}.

Despite the wide utilization of nematic liquid crystal (LC) devices, we lack an accurate theoretical model of LCs. The available models are particularly inaccurate for twisted LCs, which prevents them from entering a more extensive range of applications.
The response of LC to control voltage(s) is affected by various imperfections such as alignment layers dragging, multiple reflections, and inhomogeneity-induced depolarization. These effects represent a serious setback to the modeling of the LC response and, particularly, to the inverse task of finding the optimum control voltages to prepare the target polarization state. The theoretical model of twisted LCs \cite{YarivYeh1984} was modified to include the boundary effects \cite{Coy1996} and yet further adjusted \cite{Marquez2000,Marquez2001,Yamauchi2005} to achieve better results. Unfortunately, even with these improvements, the theoretical model of polarization transformation is not sufficiently accurate. The inaccuracy is especially apparent when considering a device consisting of multiple LC cells \cite{Zhuang1999,Bielak2021}. When aiming for a discrete set of polarization states, such a complex LC device can be calibrated and even used as a highly accurate polarimeter \cite{Bielak2021}. However, modeling of a continuous polarization response of an LC device to analog control voltages represents an open problem.

Here we model a complex twisted LC device using deep neural networks (DNNs). The model is fitted to a training dataset obtained by measuring the polarization states prepared by a device for different control voltages. We employ the mesh adaptive direct search (MADS) algorithm for black-box optimization of the deep learning model hyperparameters. The model is optimized and tested using separate datasets not involved in the training process. We demonstrate an unprecedented fidelity and repeatability between the polarization state predicted by the model and the measured state. We achieve the average infidelity of~$4 \times 10^{-4}$ of the polarization preparation for a three-cell twisted LC device. The DNN approach outperforms other models based on radial basis functions (RBF) and linear interpolation. We study the effect of the training dataset size over several orders of magnitude and found that errors of the DNN model decrease faster than the other numerical methods with an increasing number of training data. In other words, the DNN approach is more efficient with respect to the dataset size than other methods. Furthermore, we analyze the DNN model size and its overparametrization and scalability.

Our main result is solving the inverse task of finding control voltages optimal for preparing a target polarization state. We utilize the trained DNN direct model as a part of the compound autoencoder-like network to find the inverse model. This allows using physics metrics, e.g., the fidelity, consistently in the whole framework and also avoids ambiguous mapping from polarization state to control voltages. The compound model outperforms other approaches by orders of magnitude. We verify the predictive strength of the DNN model by preparing over a thousand of single-photon polarization states and performing their independent characterization using full quantum tomography. Finally, we demonstrate a remote preparation of quantum states using of entangled photons. The reported local and remote preparation of polarization-encoded quantum bits (qubits) certifies the use of twisted LC devices in quantum technology.

\begin{figure}[t]
	\centering
	\includegraphics[width=1.0\columnwidth]{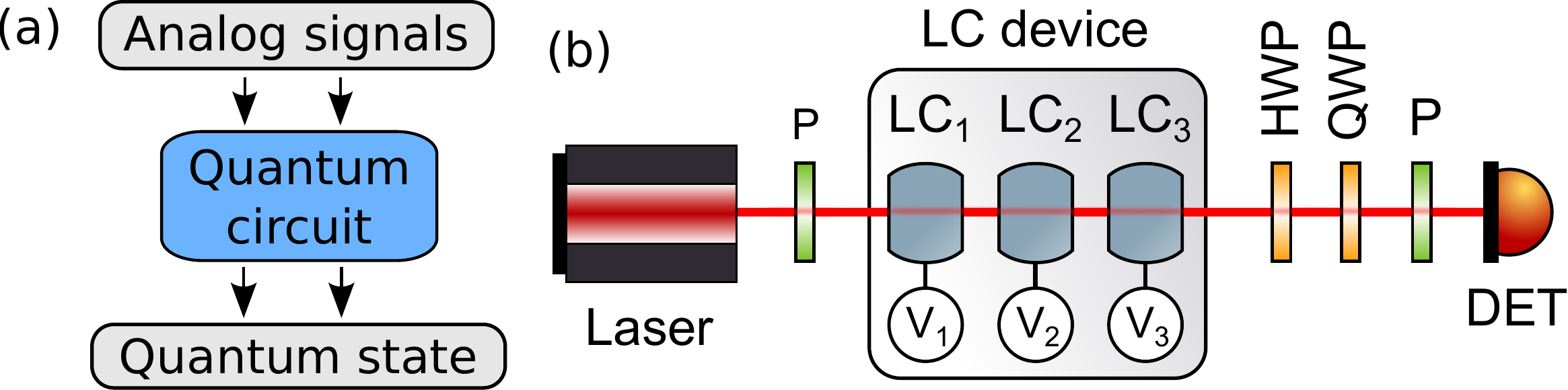}
	\caption{(a) Classical control of a quantum device: Classical analog signals modify the quantum circuit operation and affect the transformation or preparation of quantum states.
	(b) Experimental setup used to create a dataset of control voltages with corresponding polarization states: The polarizer (P) prepares a horizontal state $H$, and the liquid cells (LC${}_{1,2,3}$) induce the transformation based on applied voltages $V_1, V_2, V_3$. The prepared polarization state is analyzed using a reference polarimeter consisting of half (HWP) and quarter-wave plates (QWP), a polarizer, and a detector (DET).}
	\label{Experimental_setup}
\end{figure}

Besides the imminent application of our approach to accurate polarization qubit manipulation, we may think of it as a use case of a more general problem of optimal control of quantum devices, see Fig.~\ref{Experimental_setup}~(a). Various implementations of quantum devices share the common aspect of being controlled by classical analog signals, related non-trivially to the device operation \cite{Monroe2013, Devoret2013, Wang2019}.
The control signals need to be optimally adjusted to provide high-fidelity operation of the device \cite{Fsel2018, Cimini2019, Cimini2021, Wallnfer2020, Sgroi2021, Brown2021, Suprano2021}.
Photonic circuits on optical chips include voltage-controlled phase modulators with a complex response and crosstalk \cite{Flamini2015, Carolan2015, Jacques2019}.
Superconducting circuits are controlled by radio-frequency signals with variable amplitudes and complex timing, which are subject of optimization \cite{Werninghaus2021, Wittler2021}.
Also, semiconductor quantum dots need to be optimally tuned to produce target states \cite{Trotta2012, Plumhof2012, Baart2016, Teske2019, Zwolak2020, Moon2020, Durrer2020}.
The approach developed in this work can be directly applied to the learning of the steady-state response of quantum devices to classical control signals and, consequently, their optimal control.

\section{LC device and experimental setup}

The experimental setup implementing the polarization state transformation using liquid crystals is depicted in Fig.~\ref{Experimental_setup}~(b). Light from an 810 nm continuous-wave laser propagates through a horizontal polarizer preparing the input polarization state $H$. A device based on twisted nematic liquid crystals (LC device), described below, induces a polarization transformation on the input state controlled by an applied electric field. The prepared polarization state is characterized using full quantum tomography. The polarization state is projected into six basis states $H, V, D, A, R$, and $L$ using a reference polarimeter. It consists of half-wave and quarter-wave plates followed by a polarizer and a silicon photodiode detector. The quantum state is reconstructed from relative measurement frequencies using the maximum likelihood method \cite{Jezek2003,Hradil2004}.

The LC device consists of three independent LC cells extracted from commercially available twisted nematic displays $Lumex~LCD-S101D14TR$ by removing auxiliary layers~\cite{Bielak2021}. LCs are enclosed between glass plates with deposited electrodes producing an electric field and alignment layers forming the LC twist. The polarization transformation induced by each cell is controlled by applying voltage signals to the electrodes to manipulate the spatial orientation of LCs. We used a square wave with a 1 kHz frequency and a 50\% duty cycle, whose amplitude ranges from 0 to 10 Vpp (volts peak-to-peak). We will refer to this amplitude as a control voltage. Connecting three independent LC cells allows preparing an arbitrary polarization state by inducing the polarization transformation on a horizontal input state $H$, i.e., by applying the proper control voltages $V_1$, $V_2$, and $V_3$ to the LC device.

Using this setup, we created a dataset consisting of 27,000 combinations of three control voltages with a corresponding prepared polarization state. The average purity of these polarization states is~$(99.7-1+0.3)\%$, referring to the average value with 5th and 95th percentile. We divided the randomly shuffled dataset into three parts - training set, validation set, and test set. All discussed models were trained on the training set containing 16,000 data samples to learn the desired mapping. The validation set of 6,500 data samples was utilized for optimizing the models' hyperparameters, for example, the architecture of a neural network or a type of radial basis function. And finally, the test set containing the remaining data samples was used to evaluate the models' generalization ability on data never seen before. We note that the reference measurement induces certain infidelity to the dataset compared to the ground truth states. Therefore, as all models' predictions were evaluated compared to this dataset, the error between the ground truth states and the model predictions is a combination of the reference measurement error (independently estimated to be 99.95(5)\%) and the error of the model.

\section{Direct models and dataset-size scaling}

First, we report on predicting the prepared polarization state given the three control voltages, i.e., modeling the direct transformation. Each polarization state is described using a density matrix $\rho$. To ensure that all predictions fulfill the physical requirements on a density matrix, we utilize the Cholesky decomposition. For a Hermitian positive semidefinite matrix $M$, the decomposition has a form $M = \tau\tau^\dagger$, where $\tau$ is a lower triangular matrix with real and positive diagonal elements. Trained models output elements of the $\tau$ matrix, which are reconstructed into Hermitian and positive definite matrix $M$. This matrix is then normalized into a physically sound density matrix with unity trace, $\rho = M/$Tr$[M]$. To measure the closeness between two polarization states $\rho_{1,2}$, we use fidelity calculated as $F = ($Tr$[\sqrt{\sqrt{\rho_1} \rho_2 \sqrt{\rho_1}}])^2$, whose values range from 0 to 1, and the infidelity obtained as $1-F$.

The developed deep learning model is based on a fully connected deep neural network transforming three control voltages $V_1$, $V_2$, $V_3$ into four real-valued parameters of the $\tau$ matrix (two real diagonal and one complex off-diagonal). We performed a hyperparameter optimization of the deep learning model utilizing the MADS algorithm implemented in a black-box optimization software Nomad \cite{Nomad,Nomad_arxiv}. The optimized parameters included the number of hidden layers, number of neurons in the hidden layer, batch size, initial learning rate, and dropout regularization. The optimum dropout rate was consistently found to be zero. The optimal network consists of more than 400.000 trainable parameters arranged into 19 hidden layers with 156 neurons in each layer. We use ReLU as an activation function in each hidden layer and linear activation function in the output layer. We train the model using Adam \cite{Adam} as a stochastic gradient descend optimizer to minimize the mean squared error (MSE) loss function.
Having a significantly higher number of trainable parameters than samples in the training dataset does not pose a problem in the DNN context. Large neural networks are biased towards simpler solutions \cite{Belkin2019,Bubeck2021}, which allows using complex network architectures without overfitting.

\begin{figure}[]
	\centering
	\includegraphics[width=0.9\columnwidth]{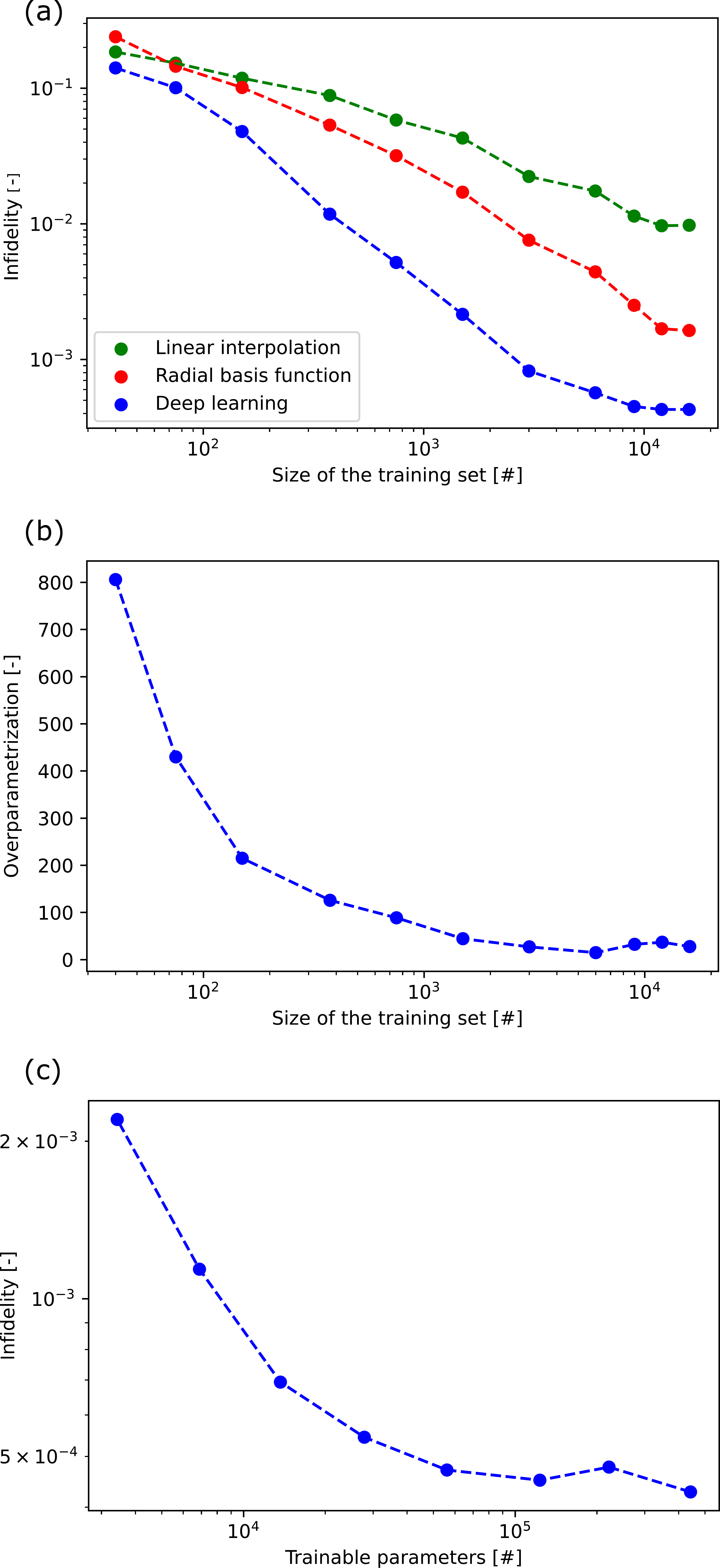}
	\caption{(a) The dependence of the optimal models' test set infidelity on the number of samples in the training set. The DNN models (blue) not only achieve the lowest errors but improve faster with the size of the training dataset than linear interpolation (green) and RBF interpolation (red). Furthermore, all three methods seem to saturate at certain infidelities for larger data sets, with the deep learning model saturating at the lowest error. (b) The ratio of the number of trainable parameters to the training set size, i.e., the overparametrization of the optimal DNNs, as a function of the number of training samples. The overparametrization decreases for large training sets. (c) The test set infidelity of sub-optimal models with the reduced number of trainable parameters. The left-most model consisting of approximately 3,500 parameters reaches the infidelity level comparable with the optimum RBF interpolation using 16,000 parameters.}
	\label{scaling}
\end{figure}

Referring to the average value with the 5th and 95th percentile, we achieve the infidelity of $(4-4-1) \times 10^{-4}$ of target polarization preparation. Here the average value lies outside of the 5th--95th percentile interval due to outliers having larger values of the infidelity.
The DNN model is by two orders of magnitude more accurate than the linear interpolation with the infidelity of $(1-1+1) \times 10^{-2}$. We also compare the DNN model to RBF interpolation, an approximation method based on a weighted sum of radial basis functions \cite{Broomhead1988}. The output variable $y_i$ is given by $\sum_j a_{ij} \varphi(\|x-x_j\|)$, where $x$ is a vector of input variables and $x_j$ is a fixed input data point termed a center. The value of a radial basis function $\varphi$ depends only on the distance from the corresponding center. A set of differently centered functions of the same type forms the basis. We tested linear, cubic, quintic, multiquadric, inverse multiquadric, and Gaussian functions. The weights $a_{ij}$ are adjusted for a minimum error on the training set. Using the validation set, we found the optimum basis function to be cubic. The resulting RBF model achieves the infidelity of $(2-2+2) \times 10^{-3}$, which is by an order of magnitude less accurate than the DNN model. The left side of Table~\ref{table_results} summarizes these results, together with the average computational time per data sample and a single CPU core. The DNN model is significantly faster in its predictions when compared to the other two methods.

Furthermore, we study the dependence of the direct models' infidelities on the number of samples in the training set, see Fig.~\ref{scaling}~\textcolor{red}{(a)}. We evaluated each point in the chart following the same procedure. First, we divided the training set into smaller disjoint subsets with the same number of samples. Each subset was used to train all three models - linear interpolation, RBF interpolation, and DNN. We used the whole validation set to optimize the hyperparameters and choose the best model for each method. The infidelities of the best models were evaluated on the test set and visualized in the chart. Each curve then represents the results of the best possible model given the number of samples in the training set. Not only does the deep learning model achieve the lowest errors and decrease faster with the size of the training set compared to both other methods, but it also saturates at errors lower by orders of magnitude, as depicted in Fig.~\ref{scaling}~\textcolor{red}{(a)}. The results can also be read that the deep learning model needs an order of magnitude lower number of experimental measurements to characterize the device with the required accuracy.

So far, we targeted the best performing DNN model using the process of hyperparameters optimization without limiting the size of the model. The optimal model lies within an overparametrized regime of the double-descent curve \cite{Belkin2019,Bubeck2021}.
The ratio of the number of trainable parameters to the number of training samples, termed overparametrization, is shown in Fig.~\ref{scaling}~(b) for the optimal DNN models trained with various numbers of the training samples.
The overparametrization reaches dozens of hundreds for very small training sets and decreases to approximately 50 for the large training sets. As seen, one can reduce the overparametrization to some extent by using larger datasets. 
Moreover, we explored sub-optimal DNNs by repeatedly lowering the number of trainable parameters of the optimal DNN by a factor of 0.5 while keeping a ratio of neurons per layer to the number of hidden layers approximately constant. The sub-optimal DNNs were trained in the same way as the optimum one except for hyperparameter optimization.
As shown in Fig.~\ref{scaling}~(c), we can significantly lower the number of trainable parameters while only slightly decreasing the accuracy.
Particularly, the sub-optimal DNN model with 13,800 trainable parameters reaches the infidelity of $(7-7+6) \times 10^{-4}$, which represents only a slight increase compared with the optimal DNN with the infidelity of $(4-4-1) \times 10^{-4}$ and more than 400 thousand parameters. We also found the smallest sub-optimal DNN performing at the same level of infidelity as the best RBF model utilizing 16,000 parameters. For this, the DNN requires only 3,500 trainable parameters. It seems that deep fully-connected networks are much more efficient and provide better scalability than interpolation approaches, including the RBF method.

\begin{table*}[ht!]
	\centering
	\begin{tabular}{|c|c|c|c|c|}
		\hline
		\multirow{2}[4]{*}{} & \multicolumn{2}{c|}{Direct model} & \multicolumn{2}{c|}{Compound model} \\
		\cline{2-5} & Infidelity & \, Time per sample \, & Infidelity & \, Time per sample \, \\
		
		\hline
		\, Linear interpolation \,  & 
		\, $(1-1+1) \times 10^{-2}$  \,    & $4 \times 10^{-3}~\textrm{s}$ &
		
		\, $(1-1+1) \times 10^{-2}$  \,   & $3 \times 10^{-3}~\textrm{s}$ \\
		
		\hline
		\, Radial basis function \,  & 
		\, $(2-2+2) \times 10^{-3}$  \,  & $4 \times 10^{-4}~\textrm{s}$ &
		
		\, $(5-5+30) \times 10^{-2}$ \,   & $5 \times 10^{-4}~\textrm{s}$ \\
		
		\hline
		\, Deep neural network \,   & 
		\, $(4-4-1) \times 10^{-4}$ \, & $8 \times 10^{-5}~\textrm{s}$ &
		
		\, $(2-2+5) \times 10^{-4}$ \, & $1 \times 10^{-4}~\textrm{s}$ \\
		
		\hline	
	\end{tabular}
	\caption{The comparison of infidelity and computational time for linear interpolation, radial basis function interpolation, and deep neural network. The infidelity values evaluated on the test dataset refer to the average, the 5th percentile, and the 95th percentile.}
	\label{table_results}
\end{table*}

\section{Inverse and compound model}

\begin{figure}[b]
	\centering
	\includegraphics[width=0.6\columnwidth]{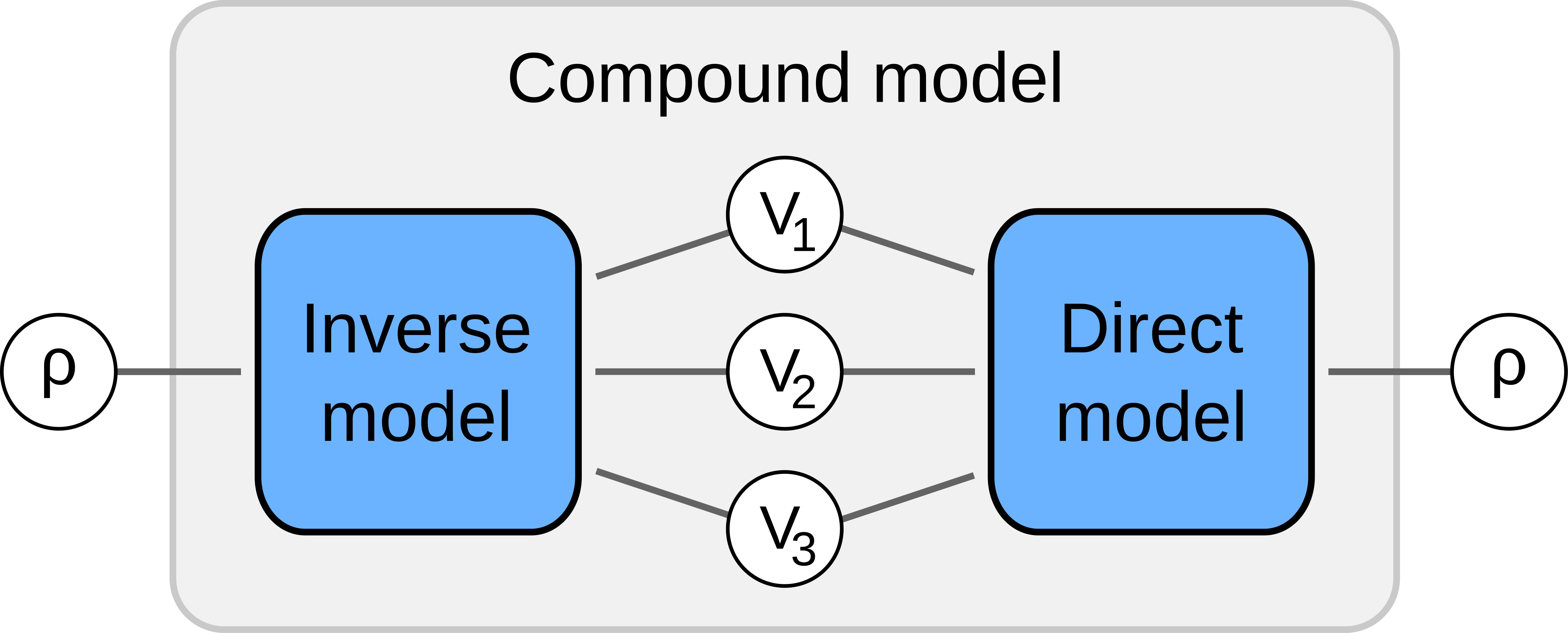}
	\caption{Representation of the compound model created by connecting the inverse model to the already-trained direct model. The model's inverse part transforms parameters of a polarization density matrix to three control voltages. The direct part then converts these voltages back into a polarization state. The direct part is fixed during the learning process, and only the inverse part is trainable.}
	\label{Compound_model}
\end{figure}

Our main goal is predicting control voltages required to prepare the target polarization state by the LC device, i.e., modeling the inverse transformation. Due to ambiguous mapping between polarization state and control voltages, the direct model cannot be easily inverted. Also, no physically fundamental/preferred metric exists in the space of the three control voltages (classical control signals in general), but the resulting inverse model would depend on the chosen metric.
To eliminate the ambiguity, we train the inverse deep learning model by connecting it to the already optimized direct model creating an autoencoder-like compound network, as visualized in Fig.~\ref{Compound_model}. Only the inverse part is trainable during the learning process, while the parameters of the direct model stay fixed. Not only does this solve the ambiguous mapping, but allows us to evaluate the performance using fidelity between the desired input state and the predicted output state. We note that the evaluated (in)fidelity describes the compound model as a whole, not just the inverse section. However, given the negligible error of the direct model, the compound model's fidelity can be viewed as an approximated accuracy of the inverse model.

The inverse section of the compound model is a fully connected DNN, which takes four elements of the polarization state density matrix for its input and predicts the three control voltages. These voltages are then transformed by the fixed direct model back into a polarization state. Utilizing MADS algorithm, we optimized hyperparameters of the inverse part of the compound model. With 91 neurons in each of the 14 hidden layers, the optimum inverse model contains more than 100.000 trainable parameters. Each hidden layer uses ReLU as an activation function, whereas the output layer utilizes the sigmoid activation function to predict control voltages rescaled to range from 0 to 1. Using Adam optimizer and MSE loss function, the optimized compound model achieves the infidelity of $(2-2+5) \times 10^{-4}$ between the input and predicted output states, evaluated on the test set.

In the same way as for DNNs, the inverse model of RBF interpolation (and linear interpolation) is a mapping from elements of the density matrix to the three control voltages. This inverse model is connected to the pretrained and optimized direct RBF model (direct linear interpolation model). The resulting RBF compound model is trained by optimizing weights of radial basis functions in the inverse model using the training data set. The training is performed for various radial functions to find the best model using the validation data set. Compared to infidelities of linear interpolation $(1-1+1) \times 10^{-2}$ and RBF interpolation $(5-5+30) \times 10^{-2}$, the deep learning model represents a significantly more accurate method to control LC devices and predict their operation. As shown on the right side of Table~\ref{table_results}, the DNN model is also the fastest in its predictions.

\section{Single-photon polarization state preparation}

\begin{figure*}[]
	\centering
	\includegraphics[width=0.85\textwidth]{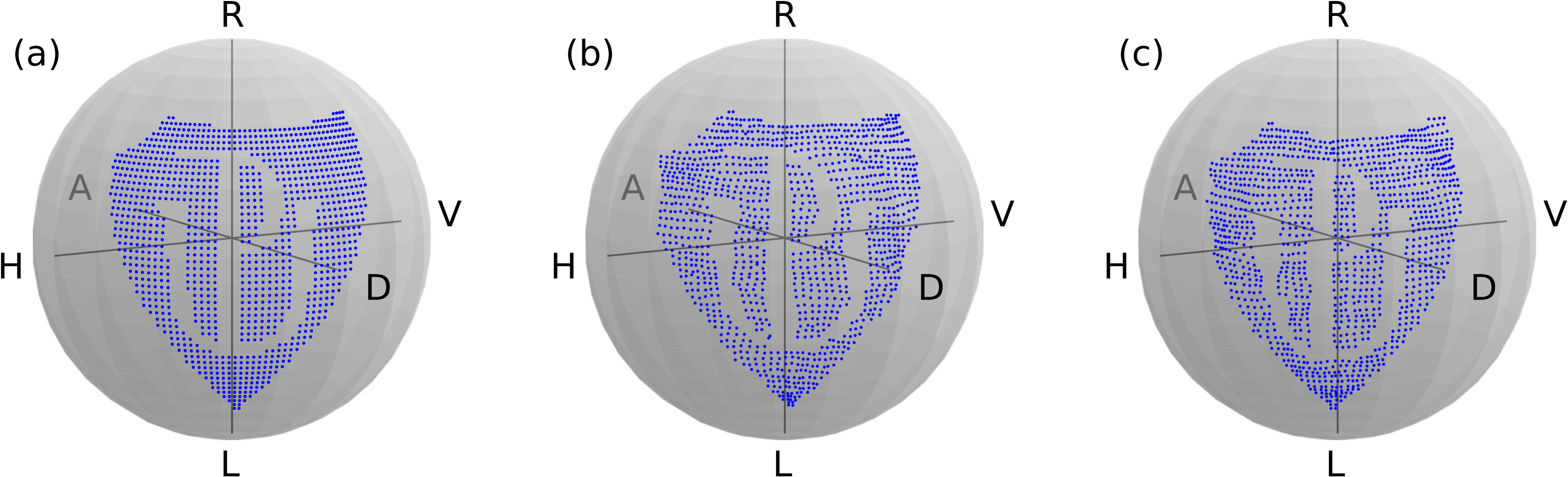}
	\caption{Visualization of the polarization states forming the Palack{\'y} University logo on the Bloch sphere. (a) The target states for which we predict the optimal control voltages using the inverse part of the compound model. (b) The logo consisting of measured polarization states from heralded single-photon preparation applying the predicted control voltages. (c) Polarization states obtained in the process of remote single-photon state preparation.}
	\label{Logo_figure}
\end{figure*}

The negligible infidelities obtained by application of the DNN models to the test dataset show unprecedentedly accurate continuous experimental control of laser beam polarization. To extend the applicability of our method even further, we implement the developed model in a single-photon polarization state preparation. Using the inverse part of our model, we predicted the control voltages for more than 1000 polarization states, with specific positions on the Bloch sphere, forming an image of the Palack{\'y} University logo, see Fig.~\ref{Logo_figure}~(a).
We prepared and fully characterized these polarization states carried by single photons.

\begin{figure}[b]
	\centering
	\includegraphics[width=1.0\columnwidth]{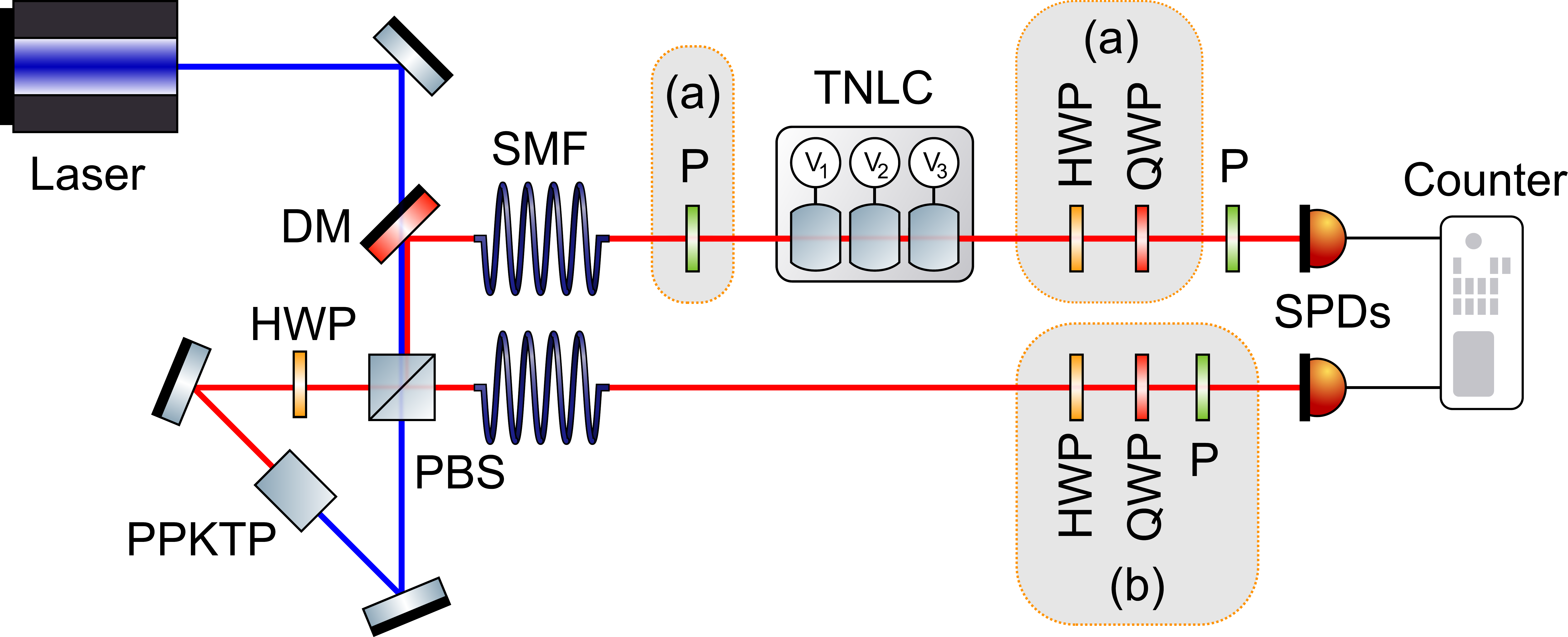}
	\caption{(a) Scheme of heralded single-photon polarization state preparation using the idler photon (lower red path) for heralding. The components denoted as (b) are not present in this configuration.
	The PPKTP crystal inside a Sagnac interferometer is pumped unidirectionally (blue line). The heralded signal photon (upper red line) in horizontal polarization state (after polarizer P) is transformed by the LC device and analyzed by quantum tomography using wave plates (HWP and QWP), a polarizer (P), and a single-photon detector (SPD). The coincidence detection events of the signal and idler detectors are acquired by a counter.
	(b) Setup for remote quantum state preparation using entangled pair of photons. The components denoted as (a) are not present in this configuration. The generation of entangled photon pairs requires bidirectional pumping of the PPKTP crystal. The signal photon is projected to a target polarization state by the LC device and the polarizer (P). The polarization state of the idler photon is analyzed by quantum tomography, see text for details.}
	\label{Logo_setup}
\end{figure}

The experimental setup is depicted in Fig.~\ref{Logo_setup} in a configuration (a). A 405-nm continuous-wave horizontally polarized laser beam propagates through a dichroic mirror (DM) and a dual-wavelength polarizing beam splitter (PBS) into a periodically poled potassium titanyl phosphate (PPKTP) crystal. In this crystal, two correlated 810-nm photons (signal and idler) are generated in the process of collinear spontaneous parametric down-conversion type-II. The signal photon (reflected by DM) is spatially and polarization filtered by a single-mode fiber (SMF) and a horizontal polarizer (P), respectively. Next, the LC device applies the target polarization transformation. A resulting polarization state of the signal photon is then analyzed using the full quantum-state tomography \cite{Jezek2003,Hradil2004}. The state is sequentially projected into six eigenstates of Pauli operators, i.e., H, V, D, A, R, and L polarizations, using half (HWP) and quarter-wave plates (QWP), polarizer, and detected by a single-photon detector (SPD). The idler photon is navigated directly to the second SPD. We measured the six projections of the signal photon in coincidence with the idler photon with a 4.5~ns coincidence window and 6~s acquisition time for each projection. The coincidence detection rate is approximately $30 \times 10^{3}$ per second. The polarization states were reconstructed using the maximum likelihood method, see Fig.~\ref{Logo_figure}~(b). The resulting logo formed by the prepared polarization states agrees very well with the one consisting of the modeled states. The average fidelity between these sets of states reaches $F = (0.998 - 0.006 + 0.002)$.

Furthermore, we demonstrate a remote state preparation~\cite{Bennett2001,Peters2005,Defienne2021}. The protocol requires a bipartite entangled state; its first qubit (signal photon) is projected to a target state (and detected) while the second qubit (idler photon) collapses to the same quantum state or a state with a local unitary operation applied. Specifically, the density matrix $\rho_{2}$ of the remotely prepared polarization state reads $\rho_{2} = \text{Tr}_1[\rho_{12} \cdot (\Pi_{1} \otimes I_2)] / \text{Tr}[\rho_{12} (\Pi_1 \otimes I_2)],$
where $\rho_{12}$ is the density matrix of the entangled state, $\Pi_{1}$ is a projector representing the LC device followed by a polarizer and the detector acting on the first subsystem, $I_2$ is the identity matrix on the second subsystem, and $\text{Tr}_1$ is a partial trace over the first subsystem. We use the entangled state close to the ideal singlet state $\rho_{12}=\ket{\Psi^{-}}\bra{\Psi^{-}}$, $\ket{\Psi^{-}} = (\ket{HV} - \ket{VH})/\sqrt{2}$, which requires the additional application of $i \, \sigma_y$ operation on the second subsystem to project the second photon into the target state.
The corresponding experimental setup is shown in Fig.~\ref{Logo_setup} as a configuration (b).
The polarization of the 405-nm laser is set to a diagonal state, and the PPKTP crystal in the Sagnac interferometer is pumped from both sides~\cite{Fedrizzi2007,Predojevi2012}. The dual-wavelength half-wave plate in the Sagnac loop swaps horizontal and vertical polarizations. The resulting two SPDC processes interfere and produce the singlet state, which is coupled to the rest of the setup by single-mode optical fibers. Unfortunately, the signal fiber is 7~m long due to space restriction in the lab, which imposes a unitary transformation randomly evolving within the measurement time. This drift would obscure the precision of the remote preparation, and an active fiber stabilization would complicate the setup considerably. Instead, we demonstrated the remote preparation numerically, including all relevant experimental imperfections. Namely, we performed the full two-qubit state reconstruction and estimated the experimental density matrix $\rho_{12}$. Its purity is $P = 0.978(1)$, concurrence $C = 0.978(1)$, and fidelity with the ideal singlet state $F = 0.987(1)$. The measurement operators $\Pi_{1}$ are obtained as output predictions of the compound model for the input polarization states forming the Palack{\'y} University logo.
The resulting states of the remote single-photon preparation are visualized in Fig.~\ref{Logo_figure}~(c). The average fidelity of these states reaches $F = (0.988 - 0.001 + 0.002)$. The remote-preparation fidelity is by $0.01$ smaller than the local-preparation fidelity, which is caused primarily by the imperfect purity of the entangled state.

\section{Conclusion}

We reported on modeling the transfer function of the complex multi-cell twisted nematic liquid crystal device using deep neural networks. The model was trained using an experimentally acquired dataset containing control voltages and tomographically measured polarization states. The trained model predicts the output polarization from the control voltages at the unprecedented fidelity level. The accuracy of the model was compared to commonly used approaches, namely linear interpolation and radial basis function interpolation. The deep learning model is more accurate and faster than both the reference methods by orders of magnitude. Also, the deep learning model is resource-efficient; it requires significantly fewer samples than other tested approaches for the given accuracy.

Our main result lies in solving the ambiguity of the control voltages in the inverse transformation. Here the optimum control voltages are predicted for a given polarization. Various combinations of control voltages can result in almost similar polarization states. Furthermore, there is no preferred metric in the space of classical control signals (control voltages in our case). We solved these issues by creating the compound model consisting of a trainable inverse part and a fixed direct part trained in the previous step. We further verified our results by employing the deep learning models in local single-photon polarization state preparation and remote quantum state preparation. Our results open the path to ultra-precise polarimetry using liquid crystals with classical light as well as with single-photon signals in quantum information processing.

Even though our work focuses on polarization encoding, we expect similar behavior and scaling in systems transforming different degrees of freedom of light such as spatial modes or which-way information in interferometric networks. The developed approach allows for near-perfect bidirectional classical control of the polarization-encoded quantum system and is easily transferable to other photonics quantum systems.

\section{Data availability}
The code and data that support the findings of this study are publicly available on GitHub~\cite{GitHub}.

\section{Funding}
The Czech Science Foundation: grant no.~21-18545S;
European Union's Horizon 2020 (2014--2020) research and innovation framework programme: QuantERA grant HYPER-U-P-S;
The Ministry of Education, Youth and Sports of the Czech Republic: grant no.~8C18002;
Palack{\'y} University: grant no.~IGA-PrF-2021-006.

\section{Acknowledgements}
We acknowledge the use of cluster computing resources provided by the Department of Optics, Palack{\'y} University Olomouc. We thank J. Provazn{\'i}k for maintaining the cluster and providing support. We also acknowledge the participation of I. Straka in developing the entanglement source.

%

\end{document}